\documentclass[prl,twocolumn,showpacs,preprintnumbers,amsmath,amssymb,nofootinbib]{revtex4}

\usepackage{amsmath}
\usepackage{amsfonts}
\usepackage{amssymb}

\newcommand{\GeV}{{\rm GeV}}

\newcommand{\eV}{{\rm eV}}

\newcommand{\be}{\begin{equation}}
\newcommand{\ee}{\end{equation}}
\newcommand{\bear}{\begin{eqnarray}}
\newcommand{\eear}{\end{eqnarray}}
\newcommand{\beqstar}{\begin{eqnarray*}}
\newcommand{\eeqstar}{\end{eqnarray*}}

\begin{document}

\preprint{HUTP-03/A006}

\title{Extranatural Inflation}

\author{Nima Arkani-Hamed, Hsin-Chia Cheng, Paolo Creminelli and Lisa Randall}
\affiliation{Jefferson Physical Laboratory, \\
Harvard University, Cambridge, MA 02138, USA}

\begin{abstract}
We present a new model of inflation in which the inflaton is 
the extra component of a gauge field in a 5d theory compactified on a
circle.
The chief merit of this model is that the potential comes only from non-local effects so that its 
flatness is not spoiled by higher dimensional operators or quantum gravity corrections. The model 
predicts a red spectrum ($n \simeq 0.96$) and a significant production of gravitational waves ($r \simeq 0.11$). 
We also comment on the relevance of this idea to quintessence.

\end{abstract}

\pacs{98.80.Cq, 11.10.Kk}

\maketitle

Inflation is a theoretically attractive idea
for solving many problems of Standard Big Bang cosmology,
and recently many experiments have confirmed its predictions of a flat
Universe with a nearly scale-invariant spectrum of adiabatic
perturbations.  However, despite its many successes, there are still no
completely natural inflationary models known in particle physics.
Although most of the literature concentrates on supersymmetric models,
supersymmetry alone is not sufficient to protect the flatness
of the inflaton potential. Supergravity corrections spoil the flatness
because supersymmetry is broken during inflation \cite{Copeland:1994vg}. 
The other candidate models for inflation rely on the inflaton taking
values larger than the Planck mass, $M_P$ ($M_P \equiv (8 \pi G)^{-1/2}$).
Examples of this possibility include  chaotic inflation \cite{Linde:gd},
and natural inflation \cite{Freese:1990rb,Adams:1992bn}, to be discussed below.
However, the use of a simple
potential requires the fine-tuning of an infinite number of
nonrenormalizable operators, suppressed by powers of $M_P$.

Natural inflation assumes that the inflaton is a Pseudo Nambu Goldstone
Boson (PNGB), parametrized by an angular variable $\theta \sim \theta + 2 \pi$.  
In the limit of exact symmetry, $\theta$ is an exactly flat direction. 
However, some tilt is necessary, and this is introduced by explicit 
symmetry breaking terms,
resulting in  a Lagrangian of the form
\begin{equation}
\label{eq:pseudo}
{\cal L} = \frac{f^2}{2} (\partial \theta)^2 - V_0 (1-\cos(\theta)) \;,
\end{equation}
where $f$ is the spontaneous breaking scale. The canonically normalized
field is $\phi = f \theta$, so the potential is naturally a function of
$\phi/f$, which can be flat for large $f$.
This scenario is however problematic, because
the requirements $\epsilon \ll 1$ and $\eta \ll 1$ on the slow-roll parameters
\begin{equation}
\label{eq:slow}
\epsilon \equiv \frac{M_P^2}{2}\left(\frac{V'}{V}\right)^2 \sim
\frac{M_P^2}{f^2} \;, \quad \eta \equiv M_P^2 \frac{V''}{V} \sim
\frac{M_P^2}{f^2}
\end{equation}
gives $f \gg M_P$. This implies the spontaneous breaking scale is
above $M_P$ which is presumably
outside  the range of validity 
of an effective field theory description. Moreover, it
is expected that quantum gravity effects, such
as the virtual appearance of black holes, will explicitly break the global symmetry \cite{Kallosh:1995hi}.
These effects, usually suppressed by powers of $f/M_P$, are here unsuppressed, so that it is hard to justify
why $V_0$ should be smaller than $M_P$, as required by the COBE bound on
the overall normalization of density perturbations: $\delta \rho/\rho \sim
10^{-5}$.

In this letter, we show that an extradimensional version of natural inflation can solve all the above
problems, giving a very simple and trustworthy model of inflation.

Consider a 5d model with the extra dimension compactified on a circle of radius $R$. The extra component
$A_5$ of an abelian gauge field propagating in the bulk cannot have a local potential, due to the higher
dimensional gauge invariance; a shift symmetry protects it similarly to what happens to a four-dimensional 
PNGB.
A non-local potential as a function of the gauge invariant Wilson loop
\begin{equation}
\label{eq:WL}
e^{i \theta} = e^{i \oint A_5 d x^5}
\end{equation}
will however be generated in presence of charged fields in the bulk. For a non-abelian
gauge group, also the self-interactions of the gauge bosons contribute to the potential. At
energies below $1/R$, $\theta$ is a 4D field with a Lagrangian of the form
\begin{equation}
{\cal L} = \frac{1}{2 \cdot g_4^2 (2 \pi R)^2} (\partial \theta)^2 - V(\theta) +
\cdots
\end{equation}
where $g_4^2 = g_5^2/(2 \pi R)$ is the 4D gauge coupling, and the potential
$V(\theta)$ is given at one-loop by
\cite{Hosotani:1983xw,Hatanaka:1998yp,Antoniadis:2001cv,vonGersdorff:2002as,Cheng:2002iz}
\begin{equation}
\label{eq:WLpot}
V(\theta) = - \frac{1}{R^4} \sum_I (-1)^{F_I} \frac{3}{64 \pi^6}
\sum_{n=1}^{\infty}
\frac{\cos(n q \theta)}{n^5} \;,
\end{equation}
where $F_I=0 (1)$ for massless bosonic (fermionic) fields of charge $q$
coupled to $A_5$. Note that the potential is of the same form as in natural inflation 
(with small corrections from additional terms in the sum), with the effective decay 
constant given by
\begin{equation}
\label{eq:decay}
f_{\rm eff} = \frac{1}{2\pi g_{\rm 4d} R} \;.
\end{equation}
It is easily seen that $f_{\rm eff}$ can be bigger than $M_P$ for sufficiently
small $g_{\rm 4d}$; the slow-roll condition $f_{\rm eff} \gg M_P$
requires only that
\begin{equation}
\label{eq:ourslow}
2 \pi g_{\rm 4d} M_P R \ll 1 \;.
\end{equation}
The canonically normalized field is $ \phi = \theta f_{\rm eff}$. Due to the higher
dimensional nature of the model, the potential (\ref{eq:WLpot}) can be trusted even
when the 4d field $\phi$ takes values above $M_P$; no dangerous
higher-dimension
operator can be generated in a local higher-dimensional theory. 
This conclusion is quite important as it is commonly believed
that any inflation model with field values above $M_P$ cannot be justified from a particle
physics point of view; we see that this conclusion is valid only if we restrict to purely
4d models. Quantum gravity corrections to the potential (\ref{eq:WLpot}) are negligible if the extra
dimension is bigger than the Planck length, different from what is expected in a 4d PNGB model. 
Again locality in the extra space is the key feature; virtual black
holes cannot spoil the gauge invariance and do not introduce a local potential
for $A_5$, while non-local effects are exponentially suppressed by $\sim e^{- 2 \pi M_5 R}$, because 
the typical length scale of quantum gravity effects (the 5d Planck length $M_5^{-1}$) is much 
smaller than the size of the extra dimension.

It is now clear that $\phi$ is a good candidate for the inflaton. The extra-dimensional nature of the
model has no cosmological consequences aside from constraining the quantum corrections to the inflaton potential,
assuming the extra dimension is stabilized.
Moreover, one can check that the Hubble length $H^{-1}$ is much larger than the size
of the extra dimension so that the theory can be treated as 4-dimensional during inflation.

To keep quantum gravity corrections under control we can estimate a lower bound on the size of the extra
dimension
\begin{equation}
\label{eq:radius}
R \gtrsim 5 \cdot M_P^{-1} \;.
\end{equation}
From this inequality the slow-roll condition (\ref{eq:ourslow}) requires
\begin{equation}
\label{eq:g4}
g_{\rm 4d} \lesssim 10^{-2} \;;
\end{equation}
this is equivalent to saying the dimensionful 5d gauge coupling satisfies
\begin{equation}
\frac{1}{g_{\rm 5d}^2} \gtrsim 10^3 M_5 \;.
\end{equation}

An upper bound on the size of the extra dimension can be obtained using the COBE normalization
for the curvature density. The maximum value of $R$ will be obtained inflating near the top of 
the potential so that the vacuum energy is given by  
\begin{equation}
\label{eq:vacuum}
V_0 \sim \frac{c}{16\pi^6}\, \frac{1}{R^4}\;, \qquad c \sim {\cal O}(1) \;,
\end{equation}
where $c$ depends on the number of charged fields in the bulk.
In this case the curvature density can be estimated by
\begin{equation}
\delta_H = \frac{1}{5\sqrt{3} \pi} \, \frac{V^{3/2}}{M_P^3 V'}\sim \frac{1}{20\sqrt{6} \pi^4}
\left( \frac{c}{M_P^4  R^4 \epsilon} \right)^{1/2} \;.
\end{equation}
The constraint of the COBE measurement
\begin{equation}
\delta_H =1.91 \times 10^{-5} \;,
\end{equation}
implies the size of extra dimension, $R$, to satisfy
\begin{equation}
\label{eq:COBEbound}
R \lesssim c^{1/4} \cdot (10^{17} \GeV)^{-1} \;,
\end{equation}
where the equality is reached taking $\epsilon$ as big as presently allowed by constraints on
$n$: $|n-1| \lesssim 0.1$. 
Note that the extra dimension is very small and it can be stabilized by a generic mechanism without
affecting the cosmological evolution up to very high scales. 

As a consequence of the smallness of the slow-roll parameters and of the density perturbations, a small 
parameter seems quite unavoidable in any model of inflation: this is the case for the gauge coupling 
(\ref{eq:g4}) in our model. Nevertheless, note that the limit $g_5 \to 0$ is natural in the 't Hooft sense. 
For $g_5 = 0$ we have a U(1) gauge symmetry with no charged field; this symmetry forbids gravity to make 
$g_5 \neq 0$ and a similar reasoning holds for non-abelian groups.

For $(2\pi g_{4d} R)^{-1} \gg M_P$, the potential can be expanded in
powers of $\phi$ and the density perturbations in the observable window is
determined by the lowest order term, the mass term. The predictions are
then the same as those of the chaotic inflation model with a parabolic potential \cite{Linde:gd}.
The spectral index is given by $n = 1-2/N_{\rm COBE}$, where $N_{\rm COBE}$
is the number of e-folds  between the exit of wavelengths now comparable to the observable Universe
and the end of inflation. The reheating temperature is determined by the $A_5$ decay through gauge interactions.
This gives $T_{\rm rh} \sim 10^{13} - 10^{14} \GeV$, when $R$ and $g_{\rm 4d}$ saturate the bounds 
(\ref{eq:radius}) and (\ref{eq:g4}) \cite{foot1} and it scales as $g_{\rm 4d}^{3/2} R^{-1/2}$. 
$N_{\rm COBE}$ can be estimated to lie in the interval $55 - 60$. The spectrum is therefore red-tilted:
\begin{equation}
\label{eq:index}
n \simeq 0.96 \;,
\end{equation}
a value not far from the present experimental sensitivity and compatible with the recent WMAP data \cite{MAP}.

It is known that a significant gravitational wave contribution requires large enough vacuum energy during inflation. 
Combined with the slow-roll conditions and the COBE normalization this requires a variation of the inflaton field
bigger than the Planck scale \cite{Lyth:1996im}, which again typically appears to be out of control of the
effective theory description.  
As we stressed, this conclusion holds only for 4d scenarios, while our model predicts a relative
amplitude of the gravitational waves and density perturbations \cite{Liddle:cg}
\begin{equation}
r \simeq 12.4 \, \epsilon = 6.2/N_{\rm COBE} \sim 0.11 \;,
\end{equation}
which could be detected by the forthcoming Planck satellite.

For $(2\pi g_{4d} R)^{-1}$ close to $M_P$, higher power terms are non-negligible and the predictions
will deviate from those based on the simple monomial potential; the
spectral index becomes in this case even smaller than (\ref{eq:index}) \cite{Moroi:2000jr} and 
therefore would be at least as measurable. On the contrary the contribution of gravity waves
becomes smaller and difficult to measure for small $f_{\rm eff}$ \cite{Adams:1992bn}. 

One could ask whether it is possible to derive  a purely
4d theory by the recent idea of deconstructing dimensions, where the Wilson line in the extra
dimension corresponds to a 4d PNGB~\cite{Arkani-Hamed:2001ca,Hill:2000mu,Cheng:2001vd,Arkani-Hamed:2001nc}.
In this case one replaces the 5d gauge theory by a (closed) chain of 4d gauge groups,
with the adjacent gauge groups connected by the link fields, which get
nonzero VEVs and break the gauge groups down to the diagonal one.
There is one linear combination of the Nambu-Goldstone bosons not eaten by the massive gauge fields.
It remains light and corresponds to the non-local Wilson line field in the 5d case.
However the required symmetry breaking scale,
\begin{equation}
f_{\rm eff} = \frac{f_{\rm link}}{\sqrt{N}} \;,
\end{equation}
where $f_{\rm link}$ is the VEV of the link fields and $N$ is the number of
the sites, requires the problematic relation $f_{\rm link} \gg M_P$. 
The point is that although
we can simulate the locality in extra dimensions in the gauge sector by deconstruction, we did not
modify the nature of the 4d gravity which is the origin of the problems \cite{foot2}. Purely 4d models
based on a PNGB with $f \ll M_P$ can however be constructed, though they involve more structure than
the simple extra-dimensional scenario we have described \cite{Arkani-Hamed:2003mz}.

Recent observations indicate that most of the energy of the Universe is
given by a negative pressure component. A candidate for this
component is a nearly massless, slow-rolling scalar, called the quintessence
field. The extreme flatness of the potential and the bounds coming from
the absence of long range forces mediated by this scalar indicate that the
quintessence field could be a PNGB \cite{Frieman:1995pm,Carroll:1998zi}.
The problems of this proposal include those of the `natural inflation'
scenario because the spontaneous breaking scale is again required to be
comparable or bigger than $M_P$ \cite{Hill:2002kq}. Again using the extra
component of a gauge field as quintessence, one can avoid this problem
and obtain a naturally flat potential.  Still, the required flatness
demands very small parameters: the quintessence mass must be smaller than
the present Hubble scale $m \sim g_{\rm 4d}/R \lesssim H_0 \simeq 10^{-33}
\eV$. Either a very small gauge coupling or a very large extra dimension
is required, in the absence of other model-building ideas.

It is interesting to ask whether the slow-roll condition $g_{\rm 4d} R M_P
\ll 1$ can naturally arise in string compactifications containing a circle
around which we can wrap a Wilson line $e^{i \theta}$. For instance if the
gauge group lives in ten dimensions and spacetime is compactified to four dimensions,
then in type I theory $g_{\rm 4d}^2 \sim g_s/(V_6 M_s^6)$ and $M_P^2 \sim g_s^{-2} M_s^8 V_6$, so
that $g_{\rm 4d} M_P R \sim g_s^{-1/2} R M_s$. For this to be much smaller than one
in the perturbative regime, we then require the radius $R \ll M_s^{-1}$, much 
{\it smaller} than the string length, which is not in the regime of validity of our effective field
theory description. Nevertheless we can take the $T$ dual to get a
convenient description of the physics. The $T$ dual theory has a radius
$\tilde{R} = 1/(M_s^2 R) \gg M_s^{-1}$, and the Wilson line for a non-abelian 
group turns into an angle $\theta$ between D-branes on the circle.  Using the distance
between branes as an inflaton is the idea of brane inflation
\cite{Dvali:1998pa}. Note however that in this regime the potential
between the branes is naturally a function of the distance $\tilde{R}
\theta$ between them, so on dimensional grounds the potential is of the
form $V(\theta) \sim M_s^4 F(\tilde{R} \theta M_s)$. Since $\tilde R M_s
\gg 1$, the largeness of the effective $f$ does not in itself guarantee
that the potential is sufficiently flat. Finding stringy scenarios that
naturally lead to $g_{\rm 4d} M_P R \ll 1$ while leading to the effective
5-dimensional field theory regime we have been using then remains an
important and interesting challenge \cite{Banks:2003sx}.

In conclusion we have shown that the extra component of a gauge field in a
5d theory is a good candidate for the inflaton. The predictions
of our model are similar to the chaotic model with parabolic potential, so
that our proposal can be considered a sensible particle physics embedding of this
simple scenario. Locality in the extra
dimension protects the flatness of the potential against Planck scale
effects, even if the inflaton takes values above the Planck scale. As there
is no trustworthy model in 4d with a variation of the inflaton field bigger than $M_P$, 
the detection of a gravitational wave contribution to the CMBR anisotropy, would
probably point to the existence of extra dimensions or other modifications of 4d gravity.


\end{document}